\documentclass[prA,twocolumn,floatfix,amsmath]{revtex4}

\usepackage{amsthm}
\usepackage{bbm}
\usepackage{amssymb}
\usepackage{verbatim}
\usepackage{graphicx}
\usepackage{epsfig}
\usepackage{slashbox}

\newtheorem{thm}{Theorem}
\newtheorem*{thm*}{Theorem}
\newtheorem{prop}[thm]{Proposition}

\newtheorem{lem}[thm]{Lemma}
\theoremstyle{definition}
\newtheorem{defn}[thm]{Definition}
\newtheorem{prope}[thm]{Property}

\newtheorem{ex}[thm]{Example}

\theoremstyle{remark}

\newcommand{\lra}{\longrightarrow}

\newcommand{\be}{\begin{equation}}
\newcommand{\ee}{\end{equation}}
\newcommand{\bea}{\begin{eqnarray}}
\newcommand{\eea}{\end{eqnarray}}
\newcommand{\1}{\mathbbm{1}}        
\newcommand{\tr}[1]{{\rm tr}\left[#1\right]}
\newcommand{\C}{\mathbb{C}}
\newcommand{\E}{\mathbb{E}}

\newcommand{\R}{\mathbb{R}}
\renewcommand{\>}{\rangle}
\newcommand{\<}{\langle}
\newcommand{\kerne}[1]{{\rm ker}\left[#1\right]}

\DeclareMathOperator{\trace}{tr}

\begin{document}

\title{Matrix Product States: Symmetries and two-body Hamiltonians}

\author{M. Sanz$^1$}
\author{M. M. Wolf$^2$}
\author{D. P\'erez-Garc\'ia$^3$}
\author{J. I. Cirac$^1$}
  \affiliation{
  $^1$Max-Planck-Institut f\"ur Quantenoptik, Hans-Kopfermann-Str. 1, 85748 Garching, Germany \\
  $^2$Niels Bohr Institute, Blegdamsvej 17, 2100 Copenhagen, Denmark \\
  $^3$Facultad de Matematicas UCM, Plaza de Ciencias 3, 28040 Madrid, Spain}

\begin{abstract}
We characterize the conditions under which a
translationally invariant matrix product state (MPS) is invariant
under local transformations. This allows us to relate the symmetry
group of a given state to the symmetry group of a simple
tensor. We exploit this result in order to prove and extend a
version of the Lieb-Schultz-Mattis theorem, one of the basic
results in many-body physics, in the context of MPS. We illustrate
the results with an exhaustive search of
 $SU(2)$--invariant two-body Hamiltonians which have such
MPS as exact ground states or excitations.
\end{abstract}


\maketitle


\begin{section}{Introduction}

Matrix Product States (MPS) \cite{FNW92,PVWC07} encapsulate many of
the physical properties of quantum spin chains. Of particular
interest in various physical contexts is the subset of
translationally invariant (TI) MPS, originally introduced as
finitely correlated states \cite{FNW92}. Their importance stems from
the fact that with a simple tensor, $A$, one can fully describe
relevant states of $N$ spins, which, at least in principle, should
require to deal with an exponential number of parameters when
written in a basis in the corresponding Hilbert space ${\cal
H}^{\otimes N}$. Thus, all the physical properties of such states
are contained in $A$. It is therefore important to obtain methods to
extract the physical properties directly from such a tensor, without
having to resort to ${\cal H}^{\otimes N}$.

An important physical property of a TI state, $\Psi$, is the
symmetry group under which it is invariant. That is, the group $G$
such that \begin{equation}
\label{eq:invariance} u_g^{\otimes N}|\Psi\rangle = e^{i\theta_g}|\Psi\rangle,
\end{equation}
where $g\in G$ and $u_g$ is a unitary representation on $\cal H$. In
a recent paper \cite{PWSVC08} we showed that for certain kind of MPS
(those fulfilling the so--called injectivity condition
\cite{FNW92,PVWC07}), this symmetry group is uniquely determined by
the symmetry group of $A$ (with a tensor product representation).
Roughly speaking this means that by studying the symmetries of $A$
we can obtain those for the whole state $\Psi$. This result allows
us, for example, to shed a new perspective into string order
\cite{PWSVC08}, a key concept in strongly correlated states in
many--body quantum systems.

Another relevant property of MPS is that they are all exact ground
states of short--range interacting (frustration free) Hamiltonians
\cite{FNW92,PVWC07}. In particular, for every TIMPS we can always
build a (so--called) 'parent' Hamiltonian for which it is the ground
state. Of particular interests are TIMPS with two--body parent
Hamiltonians; that is, whose parent Hamiltonian consist of two--body
interactions only. And among those, the ones which have a large
symmetry group, like $SU(2)$. The reason is that those are the ones
that naturally appear in condensed matter problems. Two prominent
examples are the AKLT \cite{AKLT88} and the Majumdar-Gosh
\cite{MG68} states, who have two-body parent Hamiltonians with
$SU(2)$ symmetry. They have served as toy models to understand
certain physical behavior in real physical systems, like the
existence of a Haldane gap \cite{Haldane83} in spin chains with
integer spin, or the phenomenon of dimerization \cite{MG68},
respectively. Despite their key role in the understanding of spin
chains, there are very few other examples known of TIMPS with
$SU(2)$ symmetry and with a two-body parent Hamiltonian
\cite{FNW92,KSZ93,KM07}.

In this work we first generalize the results of Ref. \cite{PWSVC08}
to arbitrary TIMPS. This enables us to derive some generic
properties about those states, as well as to obtain a simple proof
for a version of the Lieb-Schultz-Mattis theorem \cite{LSM61}. This
celebrated theorem states that all Hamiltonians with $SU(2)$
symmetry are gapless for semi-integer spin (dim$({\cal H})=n+1/2$,
$n=0,1,\ldots$). In our case, we can prove that all TIMPS
corresponding to systems with semi-integer spins cannot be the
unique ground state of a local frustration-free Hamiltonian.
Furthermore, we can extend the proof to other groups, like $U(1)$
for spin 1/2 systems, and find counterexamples for this last case
when the spin is 5/2 or larger.

In the second part of our work we concentrate on MPS that are
eigenstates (not necessarily grounds states) of  a (so--called
'parent') Hamiltonian which has $SU(2)$ symmetry and contains
two--body interactions only. We find other families of Hamiltonians
beyond the well--known AKLT and Majumdar-Gosh with those features.
Furthermore, we find the first examples of MPS that correspond to
excited states of $SU(2)$-invariant Hamiltonians. There is a new example of state with spin $1$, which
is never the ground state of any frustration free $SU(2)$-invariant two-body
hamiltonian. In order to make a systematic search of all those MPS
we develop a simple technique that allows for a numerical systematic
search.


This paper is organized as follows. In Section \ref{Sec:pre} we
review some of the basic properties of TIMPS and establish the
notation that will be needed in the following. In Section \ref{2} we
establish the relation between the symmetry group of a TIMPS and
that of the tensor $A$ defining the MPS. For continuous symmetries,
such as $SU(2)$, we will see that the set of symmetric TIMPS is
intimately related to the set of Clebsch-Gordan coefficients.
Section \ref{Sec:LSM} then provides an MPS version of the
Lieb-Schultz-Mattis theorem and in Section \ref{Sec:TBH} we give a
detailed investigation of $SU(2)$ symmetric TIMPS which are
eigenstates of two-body Hamiltonians.
\end{section}

\begin{section}{Matrix Product States}\label{Sec:pre}
Let us consider a system with periodic boundary conditions of $N$
(large but finite) sites, each of them with an associated
$d$-dimensional Hilbert space. A translationally invariant MPS on
this system can be defined with a valence bond construction in the
following way: Let us consider another couple of $D$ dimensional
ancillary/virtual Hilbert spaces associated to each site and
connected to the real/physical $d$ dimensional space by a map
$\mathcal{A} =  \sum_{i \, \alpha \, \beta} A_{i , \alpha \beta}
|i\> \< \alpha \beta |$. Then, by introducing maximally entangled
states connecting every pair of neighboring virtual Hilbert spaces
(usually called entangled \textit{bonds}), it is not difficult to
prove that the state can be written as
\begin{equation*}
|\Phi \> = \sum_{i_1,\ldots, i_N} \tr{A_{i_1} \cdots A_{i_N}} |i_1 \cdots i_N \>
\end{equation*}
where we call the matrices $\mathcal{K} = \{ A_i \in \mathcal{M}_D
\, , \, i= 1, \ldots, d \}$  \textit{Kraus operators}. A way to work
simultaneously with all of them is to define the map
\begin{equation}\label{IsometryKraus}
V = \sum_i A_i \otimes |i\>\;.
\end{equation}

\begin{figure}[h!]\label{MPS}
 \begin{center}
 \includegraphics[width=0.5\textwidth]{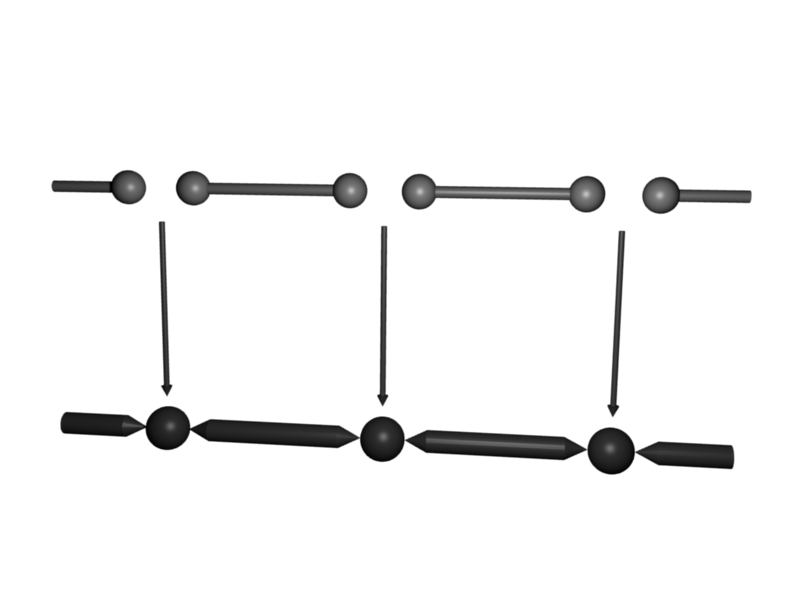}
  \caption[Fig. 1]{\footnotesize \emph{This figure represents the
  MPS construction. A
  pair of virtual spins which are connected to their neighbors via a maximally entangled state $|\Omega > = \sum_{\alpha = 1}^d |\alpha \; \alpha >$ are mapped into the physical spins (below).
  All properties of the state originate from the mapping between physical and virtual system.}}
 \end{center}
\end{figure}

For each MPS there exists a canonical form \cite[Theorem III.7,
Lemma IV.4]{PVWC07} which assures that one may choose all matrices
$A_i$ with a block diagonal structure \footnote{Eventually after
gathering some spins together to neglect the periodic
components---something that we will always assume.}, in such a way
that after gathering enough spins together, the Kraus operators
fulfil:

\begin{prope}[\it{Span property}]\label{p*}
The set of products $\mathcal{P} = \{ A_{i_1} \cdots A_{i_n} \}$,
with $n$ the collected spins, spans the vector space of all matrices
with the same block diagonal structure.
\end{prope}

It is an open conjecture stated in \cite{PVWC07} and verified in
many particular cases, that an upper bound for the number of sites
which have to be gathered to achieve property \ref{p*} depends
only on the dimension $D$ of the Kraus operators. When there is only
one block in the above \emph{canonical decomposition} the MPS is
usually called {\it injective}, since the linear operator mapping
boundary conditions to the resulting states is indeed injective
\cite{FNW92, PVWC07} when taking sufficiently many particles. The
definition reads:
\begin{prope}[\it{Injectivity}]\label{}
There exists $n$ such that the map
$\Gamma_n(X)=\sum_{i_1,\ldots,i_n}\trace(XA_{i_1}\cdots
A_{i_n})|i_1\cdots i_n\>$ is injective.
\end{prope}

For each MPS $|\psi\>$ one can construct a Hamiltonian, called \textit{parent
Hamiltonian}, for which $|\psi\>$ is an eigenstate with eigenvalue $0$ .
\begin{defn}[\it{Parent Hamiltonian}]\label{parham}
Let $\rho^{(k)}$ be the reduced density matrix  of $|\psi\>$ for $k$
particles ($k$ will be called the {\it interaction length} of the
parent Hamiltonian). Let us suppose that $\{ |v_i \rangle
\}_{i=1}^{r}$, with $r\ge 1$, is an orthonormal basis for
$\kerne{\rho^{(k)}}$. Taking any  linear combination of projectors
$h (\vec{a})= \sum_{i=1}^r a_i |v_i \rangle \langle v_i |$, we
define $H = \sum_i \tau_i (h)\otimes \1_{\rm rest}$, where $\tau_i$
is the translation operator.
\end{defn}

If $a_i \ge 0$, then the Hamiltonian is  positive semidefinite and
$|\psi\>$ is indeed a ground state. Moreover $H$ is {\it frustration
free}, since $|\psi\>$ minimizes the energy locally. Injectivity has
now a deep physical significance. If it is reached for $n$ particles
and every $a_i>0$, it ensures that the MPS is the {\it only} ground
state of its $(n+1)$-local parent Hamiltonian, that it is an
exponentially clustering state and that there is a gap above the
ground state energy \cite{FNW92,PVWC07}.

In this work we will focus on symmetries of {\it states} instead of
{\it Hamiltonians}. There is however a close connection between the
two approaches. On the one hand, it is clear that the unique ground
state of a symmetric Hamiltonian has to keep the symmetry. On the
other hand, we have the following
\begin{prop}
If an MPS $|\psi\>$ is invariant under a representation of a group,
one can choose its parent Hamiltonian $H$ invariant under the same
representation.
\end{prop}
To see that it is enough to notice that the symmetry in the state
(\ref{eq:invariance}) implies the invariance of $\kerne{\rho^{(k)}}$
under the same symmetry. Symmetrizing $\kerne{\rho^{(k)}}$ (i.e.,
averaging it) w.r.t. the considered group will then yield a
symmetric $h$ which still constitutes a parent Hamiltonian.
\end{section}


\begin{section}{Locally symmetric MPS}\label{2}

In this section we analyze the implications of a given symmetry for
a MPS. First, we  show that the symmetry transfers to the Kraus
operators---generalizing the findings of \cite{FNW92,PWSVC08}. In a
second step we  show that the symmetry in the Kraus operators
imposes that they are essentially uniquely defined in terms of
Clebsch-Gordan coefficients. Finally, for the special case of
$SU(2)$ one can simplify even further and analyze the qualitative
differences between integer and semi-integer spin.


\begin{subsection}{Characterization of symmetries}
It was demonstrated in \cite{PWSVC08} that the Kraus operators which
describe any {\it injective} state symmetric under a group $G$ fulfil the
condition $\sum_i u_{i j}^g A_i = U_g A_j U_g^{\dagger}$, where $u$
and $U$ are representations of $G$. We provide in this section a
generalization in which injectivity is not required.
The $N$ appearing in the proof must be sufficiently large to obtain property
\ref{p*} after collecting $N/5$ spins.

We start by proving the case of discrete symmetries, extending the
demonstration to continuous groups below.

\begin{thm}[Discrete symmetries]\label{thm:MPS-sym}
Let $\{A_i\}_{i=1}^d$ be the Kraus operators which describe a
locally invariant MPS $|\psi\>$ with respect to a single unitary
$u$, i.e. $u^{\otimes N}|\psi\>=e^{i\theta}|\psi\>$. Then, the
symmetry in the physical level can be replaced by a local
transformation in the virtual level. This means that there exists a
unitary $U$ -- which can be taken block diagonal with the same
block structure as the $A$'s in the MPS and composed with a
permutation matrix among blocks, i.e. $U = P(\oplus_b V_b)$ -- such
that
\begin{equation}\label{eq:MPS-sym}
\sum_{j} u_{ij} A_j=W UA_iU^\dagger
\end{equation}
with $W=\oplus_be^{i\theta_b}\1_b$.
\end{thm}

\begin{proof}
We follow here a reasoning as in the proof of \cite[Lemma
IV.4]{PVWC07}. We collect the spins in five different blocks, each
one of them with property \ref{p*}. Applying $u^{\otimes N}$ gives
us the same MPS (we incorporate the global phase in the new
matrices) with different matrices $B$'s, but with the same block
diagonal form and also (after gathering) with property \ref{p*}. We
now require the following lemma, which is demonstrated below.

\begin{lem}
For each block in the $A$'s, for instance the one given by matrices
$A^1_i$, there is a block in the $B$'s, given by matrices $B^1_i$,
which expands the same MPS.
\end{lem}

Since both are now canonical forms of the same injective MPS, by
\cite[Theorem 3.11]{PVWC07}, \footnote{The condition of
\cite[Theorem 3.11]{PVWC07} that the canonical form in the OBC must
be unique can be dropped by eq. (3) of [Lamata et al,
PRL 101, 180506 (2008)]}, they must be related by a unitary and a phase:
$V_1 A^1_i V_1^\dagger=e^{i\theta_1}B^1_i$, which finishes the proof
of the theorem.

Let us prove now the lemma. By using property \ref{p*} and summing
with appropriate coefficients, it is possible to show that there
exists a block diagonal $D\times D$ matrix $X\not = 0$ such that

\begin{equation*}
\tr{A^1_{i_2}\cdots
A^1_{i_5}}=\tr{X B_{i_2}\cdots B_{i_5}},\quad \forall i_2,\ldots,
i_5
\end{equation*}

Since $X\not =0$, there exists one block, let us say $X_1$, different from $0$. Then, summing with
appropriate coefficients again we get that there exists a matrix
$Y\not =0$ such that
\begin{equation*}
\tr{Y A^1_{i_3}A^1_{i_4}A^1_{i_5}}=\tr{X_1 B^1_{i_3}B^1_{i_4} B^1_{i_5}},
\quad \forall i_3,i_4, i_5
\end{equation*}
We can now argue as in \cite[Lemma
IV.4]{PVWC07} to conclude the proof.
\end{proof}

If we have now a symmetry given by a compact connected Lie group
$G$, that is, (\ref{eq:invariance}) holds for any $g\in G$ and a
representation $g\mapsto u_g$, we obtain the following.

\begin{thm}[Continuous symmetries]\label{thm:G-sym}
The map $g\mapsto P_g$ is a representation of $G$ and therefore the
trivial one. The maps  $g\mapsto e^{i\theta^b_g}$ and $g\mapsto
V_g^b$ are also representations of $G$.
\end{thm}

\begin{proof}
Let us start with the map $g\mapsto P_g$. From eq.
$\eqref{eq:MPS-sym}$ we get
\begin{equation}\label{eq:2}
\begin{split}
W_{g_2g_1}U_{g_2g_1}A_hU_{g_2g_1}^{\dagger} & =\sum_j u_{jh}^{g_2g_1}A_j =\\
\sum_{jk} u_{jk}^{g_2}u_{kh}^{g_1} A_j & = W_{g_2}W_{g_1,P_{g_2}}
U_{g_2}  U_{g_1} A_h U_{g_1}^{\dagger} U_{g_2}^{\dagger}
\end{split}
\end{equation}
where $W_{g_1,P_{g_2}}$ is the same unitary as $W_{g_1}$ but with
the blocks permuted according to the permutation $P_{g_2}$. Since
$P_{g'}W_g=W_{g,P_{g'}}P_{g'}$ and $W_g$ commutes with all other
terms appearing in eq. (\ref{eq:2}), we can multiply successively
and use property \ref{p*} (with $L$ the required block size), to
get, for all $n\ge L$ and all $X$ block-diagonal,
\begin{equation}\label{eq:2-prime}
W_{g_2g_1}^nU_{g_2g_1}XU_{g_2g_1}^{\dagger} = (W_{g_2}
W_{g_1,P_{g_2}})^nU_{g_2}  U_{g_1} X U_{g_1}^{\dagger} U_{g_2}^{\dagger}\; .
\end{equation}
By taking $X=\1_b$ for each block $b$,
we get that $P_{g_2}P_{g_1}$ must be $P_{g_2g_1}$. But since we are
assuming the group $G$ connected, this in turn
implies that $P_g=\1$ for all $g$. With this we can split equation
(\ref{eq:2-prime}) into blocks to get, for each $b$, each $n\ge L$
and each matrix $X$,

\begin{equation}\label{eq:2-prime-prime}
e^{in\theta^b_{g_2g_1}}V^b_{g_2g_1}XV^{b\;\dagger}_{g_2g_1} =
e^{in(\theta^b_{g_1}+\theta^b_{g_2})}V^b_{g_2}  V^b_{g_1} X
V_{g_1}^{b\;\dagger} V_{g_2}^{b\;\dagger}
\end{equation}
 Taking $X=\1$ we obtain
\begin{equation*}
e^{i n (\theta^b_{g_2 g_1})} = e^{in(\theta^b_{g_1} + \theta^b_{g_2})}
\end{equation*}
In particular, when $n = L$, we get that $L(\theta^b_{g_2 g_1}) =
L(\theta^b_{g_1} + \theta^b_{g_2})+2 k_0 \pi$ and when $n = L + 1$
that $(L+1)(\theta^b_{g_2g_1}) = (L+1)(\theta^b_{g_1} +
\theta^b_{g_2}) + 2 k_1 \pi$. Gathering both results, the $L$ can be
removed and we obtain
$\theta^b_{g_2 g_1} = \theta^b_{g_1} + \theta^b_{g_2} + 2(k_1 - k_0)\pi$.\\

Finally, to show that $g\mapsto V^b_g$ is a representation, it is
enough to notice that eq. (\ref{eq:2-prime-prime}) implies that
$V_{g_1}^{b\;\dagger} V_{g_2}^{b\;\dagger} V^b_{g_2 g_1}$ commutes
with every matrix.
\end{proof}

\begin{figure}[h!]\label{Krauscond}
 \begin{center}
  \includegraphics[width=0.5\textwidth]{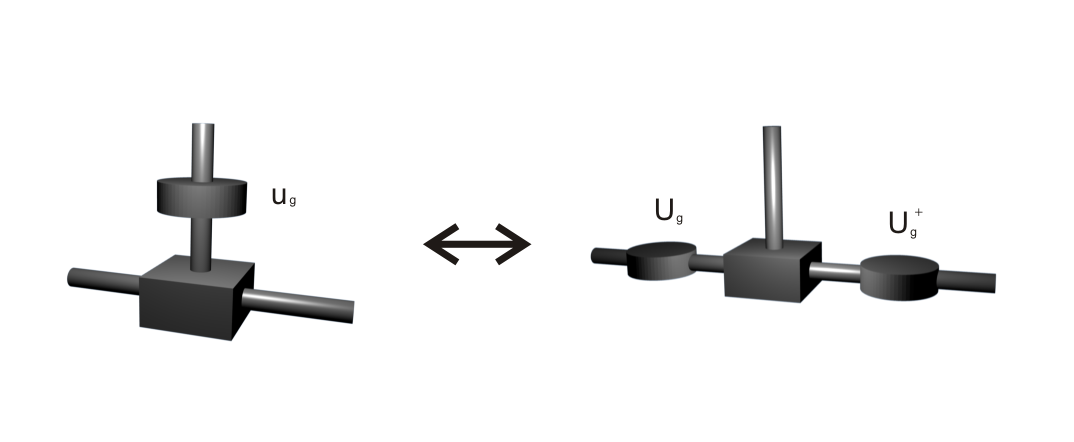}
  \caption[Fig. 2]{\footnotesize \emph{The unitary $u_g$ applied on
  the physical level is reflected in the virtual level as a pair of
  unitaries $U_g$.}}
 \end{center}
\end{figure}

A trivial consequence of  these theorems is the fact that having an
{\it irreducible} representation $U_g$  in the virtual level implies
that the MPS has to be {\it injective}. We give an alternative proof
of this fact in the appendix without having to rely on the MPS
canonical form. There we analyze also when the reverse implication
holds.
\end{subsection}


\begin{subsection}{Uniqueness of the construction method}

Once the theorem which provides the condition that the Kraus
operators must fulfil in order to generate invariant MPS has been
established, the next step is to prove that they can always be
constructed by means of Clebsch-Gordan coefficients. To do that, it
is more convenient to work with the map $V$ defined in
(\ref{IsometryKraus}). From the definition it is clear that the
condition $\sum_{i}u_{ij}^gA_i=U_gA_jU_g^\dagger$ reads then
$U_g\otimes u_gV=VU_g$. Notice that we have removed the dependence
on the phase. By Theorem \ref{thm:G-sym} this can be done for groups
with a complex enough structure, as $SU(2)$, for which there is no
non-trivial one-dimensional representation.

Given a compact group $G$, the tensor product of two irreps --we are
choosing a single representative for each class of equivalent
irreps-- can always be decomposed as a direct sum of irreps
\begin{equation*}
   u_g \otimes v_g C = C \bigoplus_i c^{i}_g
\end{equation*}
where $C$ is a unitary whose elements are called Clebsch-Gordan
coefficients. In what follows we will denote by
$\phi_i:\C^{d_i}\rightarrow \C^d\otimes \C^{d'}$  the matrix
associated to the restriction of $C$ to the  $d_i$-dimensional
invariant subspace $\mathcal{H}_i$ associated to the irrep $c^i_g$,
with $d,d'$ being the dimensions of the representations $u_g$ and
$v_g$ respectively.

We are interested in possible solutions of
\begin{equation} \label{eq1}
  u_g \otimes v_g \Omega = \Omega w_g\quad \forall g\; .
\end{equation}
where $u_g,v_g,w_g$ are irreps of a given compact group $G$. It is
clear that taking
\begin{equation}\label{eq:omega}
  \Omega= \sum_i \beta_i \phi_i
\end{equation}
does the job if we sum over $i$'s corresponding to equivalent
representations $c^i_g=w_g$. The next lemma guarantees that this is
all.

\begin{lem}\label{lem:C-G}
All possible solutions of Equation (\ref{eq1}) are given by (\ref{eq:omega}).
\end{lem}

\begin{proof}
Any $\Omega$ verifying eq. $\eqref{eq1}$ gives
\[
   \Omega^{\dagger} \Omega = w_g \Omega^{\dagger} \Omega w_g^\dagger
\]
which means by Schur's lemma that $\Omega^{\dagger} \Omega = \alpha
\mathbb{I}$ and we may assume that, if there is a non-zero solution,
it can be taken an isometry. Moreover, introducing
$V=C^\dagger\Omega$, which verifies $V^\dagger V=\mathbb{I}$, one
has
\begin{equation}\label{eq:lema-clebsch}
Vw_g= \left(\oplus_i c^i_g\right) V
\end{equation}
From there one gets that $P=VV^\dagger$ is a rank $d$ projector ($d$
the dimension of the representation $w_g$) that commutes with
$\left(\oplus_i c^i_g\right)$ for all $g$. By Schur's lemma, it is
supported on $\oplus_i \mathcal{H}_i$ with $i$'s such that
$c^i_g=w_g$ and in this subspace it is of the form
 $$\left(
     \begin{array}{ccc}
       |\beta_1|^2 \1_d & \bar{\beta_1}\beta_2\1_d & \cdots \\
       \beta_1\bar{\beta_2}\1_d & |\beta_2|^2\1_d & \cdots \\
       \cdots & \cdots & \cdots \\
     \end{array}
   \right)=|\beta\>\<\beta|\otimes \1_d\; .
 $$
This implies that $V=|\beta\>\otimes W$ for a given $d\times d$
unitary $W$. But if we substitute this in (\ref{eq:lema-clebsch}),
since we are assuming a unique fixed representative for each class
of equivalent representations, we get $W=\1_d$ and $\Omega=\sum_i
\beta_i\phi_i$.
\end{proof}

From this we can now conclude:

\begin{thm}\label{thm:cl-go}
Let us consider a group $G$ and two representations $u_g$ (irrep)
and $U_g =\bigoplus_i U^{D_i}_g$. Then, the structure of  all
possible maps $V$ fulfilling $U_g \otimes u_g V = V U_g$ is
    \begin{equation} \label{eq3}
       V =
        \begin{pmatrix}
           \alpha_{1 1} V_{D_1}^{D_1} & \alpha_{1 2} V_{D_1}^{D_2} &  \cdots & \alpha_{1 n} V_{D_1}^{D_n}\\
           \alpha_{2 1} V_{D_2}^{D_1} & \alpha_{2 2} V_{D_2}^{D_2} &  \cdots & \alpha_{2 n} V_{D_2}^{D_n}\\
           \vdots & \vdots & \ddots & \vdots \\
           \alpha_{n 1} V_{D_n}^{D_1} & \alpha_{n 2} V_{D_n}^{D_2} &  \cdots & \alpha_{n n} V_{D_n}^{D_n}
        \end{pmatrix}
    \end{equation}
where $V_{D_i}^{D_j}$ is a solution, according to Lemma
\ref{lem:C-G}, to $U^{D_i}_g \otimes u_gV_{D_i}^{D_j} =
V_{D_i}^{D_j} U^{D_i}_g$.
\end{thm}
\end{subsection}


\begin{subsection}{The case of  $SU(2)$}

Let us apply the results of the previous section to the case in
which $G = SU(2)$. Our construction is a natural generalization of
the one used in \cite{FNW92,SZV07}.

We consider from now on irreducible representations $u_g$ of the
symmetry on the physical spin. Nevertheless, a substantial part of
the results can be straightforwardly extended to the reducible case.
Hence, we are interested in analyzing the restrictions that $SU(2)$
impose in the general solution given by Theorem \ref{thm:cl-go} to
the equation
\begin{equation}\label{eq:spin-V}
(U \otimes J) V = V U
\end{equation}
where, with some abuse of notation, $J$ is the $SU(2)$ irrep
corresponding to spin $J$ \ and $U = (i_1 \oplus \ldots \oplus i_n
\oplus s_1 \oplus \ldots \oplus s_m)$ is the virtual representation
composed of $n$ integer irreps and $m$ semi-integer irreps. Note
that in the Clebsch-Gordan decomposition of $SU(2)$ all
representations appear with multiplicity one. Therefore there is
only one term in the sum in (\ref{eq:omega}). At this point one
should distinguish the cases of $J$ integer or semi-integer. If $J$
is integer, zero is the only solution to $(i_j\otimes
J)\Omega=\Omega s_k$ and $(s_k\otimes J)\Omega=\Omega i_j$ for all
$j,k$, and we get in (\ref{eq3}) a block diagonal structure:
\begin{equation*}
  V  =
        \begin{pmatrix}
           \alpha_{1}^1 V_{i_1}^{i_1} &  \cdots & \alpha_{1}^n V_{i_1}^{i_n} & 0 & \cdots & 0 \\
           \vdots  & \ddots & \vdots  & \vdots & \ddots & \vdots \\
           \alpha_{n}^1 V_{i_n}^{i_1}  &  \cdots & \alpha_{n}^n V_{i_n}^{i_n} & 0 & \cdots & 0 \\
           0 & \cdots & 0 & \alpha_{n+1}^{n+1} V_{s_1}^{s_1} &  \cdots & \alpha_{n+1}^{n+m} V_{s_1}^{s_m} \\
           \vdots  & \ddots & \vdots & \vdots & \ddots & \vdots \\
           0 & \cdots & 0 & \alpha_{n+m}^{n+1} V_{s_m}^{s_1} &  \cdots & \alpha_{n+m}^{n+m} V_{s_m}^{s_m} \\
        \end{pmatrix}
\end{equation*}

The paradigmatic example in this case is the {\it AKLT} state
\cite{AKLT88}, which corresponds to the case of $J=1$, $U=1/2$ in
(\ref{eq:spin-V}). In \cite{FNW92}, the authors generalized the AKLT
model to arbitrary integer $J$ and $U$ irreducible. We will call the
resulting MPS {\it FNW states}. It is shown in \cite{FNW92} how for
$U=\frac{J}{2}$  FNW states are unique ground states of frustration
free nearest-neighbor interactions.  An alternative construction
focused on the restrictions imposed by the $SU(2)$ symmetry on the
density matrix instead of the Kraus operators can be found in
\cite{DMNS98}.

If $J$ is semi-integer, zero is the only solution to $(s_j\otimes
J)\Omega=\Omega s_k$ and $(i_k\otimes J)\Omega=\Omega i_j$ for all
$j,k$, and we get in (\ref{eq3}) an off-diagonal structure:
\begin{multline*}
  {V} = \\
        \begin{pmatrix}
           0 & \cdots & 0 & \alpha_{1}^{n+1} V_{i_1}^{s_1} &  \cdots & \small \alpha_{1}^{n+m} V_{i_1}^{s_m} \\
           \vdots  & \ddots & \vdots & \vdots & \ddots & \vdots \\
           0 & \cdots & 0 & \alpha_{n}^{n+1} V_{i_n}^{s_1} &  \cdots & \alpha_{n}^{n+m} V_{i_n}^{s_m} \\
           \alpha_{n+1}^1 V_{s_1}^{i_1} &  \cdots & \alpha_{n+1}^n V_{s_1}^{i_n} & 0 & \cdots & 0 \\
           \vdots  & \ddots & \vdots  & \vdots & \ddots & \vdots \\
           \alpha_{n+m}^1 V_{s_m}^{i_1}  &  \cdots & \alpha_{n+m}^n V_{s_m}^{i_n} & 0 & \cdots & 0 \\
        \end{pmatrix}
\end{multline*}

It is clear that the virtual representations must be reducible now,
which is very much related to the Lieb-Schultz-Mattis theorem, as we
will show in the following section. The paradigmatic example in this
case is the \textit{Majumdar-Ghosh model} \cite{MG68}, which
corresponds to $J=\frac{1}{2}$ and $U= \frac{1}{2} \oplus 0$. A
generalization of this model for the case of arbitrary $J$ and
$U=F\oplus 0$,  was recently proposed in \cite{KM07}.

In general, it is possible to find a set of representations which
fits into any model with $SU(2)$ symmetry, for instance
\cite{KSZ93,RSDM98,RG08,Ku02}.
\end{subsection}
\end{section}


\begin{section}{Lieb-Schultz-Mattis Theorem}\label{Sec:LSM}

The Lieb-Schultz-Mattis theorem states that, for semi-integer spin,
a $SU(2)$-invariant 1D {\it Hamiltonian} cannot have a uniform
(independent of the size of the system) energy gap above a unique
ground state. That is, symmetry imposes strong restrictions on the
possible behaviors of a system. In this section we want to go a step
further and analyze which implications one can obtain from having a
{\it single} symmetric {\it state} in a semi-integer spin chain. By
restricting our attention to the class of MPS we will show
\begin{thm}\label{LSM}
Any MPS with an $SU(2)$ symmetry in the sense of
(\ref{eq:invariance}) with $u_g$ irrep and even physical dimension
$d$ cannot be injective. By Theorem 11 of \cite{PVWC07} this implies
that it cannot be the unique ground state of {\emph any} frustration
free Hamiltonian.
\end{thm}

\begin{proof}
Let us assume that the MPS is injective and prove the theorem by
contradiction. Theorems \ref{thm:MPS-sym} and \ref{thm:G-sym}
guarantee that
\begin{equation}\label{eq7}
\sum_j u_{j k}^g A_j = U_g A_k U_g^{\dagger}\;.
\end{equation}
We consider $u = e^{i J_z}$ with $(J_z)_{j,k} =
\delta_{j,k}\big(k-(d+1)/2\big)$, $k=1,\ldots,d$. Then, eq.
$\eqref{eq7}$ gives
\begin{equation}\label{eq:lsm}
e^{i\;\varphi_k}
A_k = U A_k U^{\dagger}
\end{equation}
for a unitary $U$ and $\varphi_k$ half-integer. We finish by proving
that if $N$ is odd, $\trace(A_{k_1} \cdots A_{k_N}) = 0$ and hence
the MPS cannot be injective. From (\ref{eq:lsm}) we get
$\trace(A_{k_1} \cdots A_{k_N}) = 0$ unless $\sum_{i=1}^N
\varphi_{k_i}=N(d+1)/2$. The latter is, however, impossible for $N$
odd as then the l.h.s. is integer whereas the r.h.s. is
half-integer.
\end{proof}

From the proof one may get the impression that only $U(1)$ symmetry
is required, and this is indeed the case if the generator of such
symmetry has eigenvalues $-m/2,...,m/2$ as above. The next example
shows that this is, however, not true for {\it any} $U(1)$ symmetry,
which in turn shows that a larger symmetry like $SU(2)$ is required
for the Lieb-Schultz-Mattis theorem.

\begin{ex}
Let us consider a local symmetry generated by $G = e^{i \beta H}$
for a hermitian matrix $H$. Let us choose the
physical dimension $d = D^2 - D$, which is always even, and the set
of Kraus operators $\mathcal{K}= \left\lbrace A_{(i,j)} = |i
\left\rangle \right\langle  j |, \; i \neq j \right\rbrace$. Select $\alpha_1 , \ldots , \alpha_D \in \R$ such that $\alpha_i -
\alpha_j \neq 0$ if $i \neq j$ and $H$ the
diagonal matrix $H = \sum_{i \neq j} (\alpha_i - \alpha_j) |(i,j)
\rangle \langle (i,j)|$ (which has in addition only non-zero eigenvalues). With
$U_{\beta} = e^{i \beta \Omega}$ where $\Omega = diag [\alpha_1
\ldots \alpha_D]$ it is clear that
\begin{equation*}
e^{i \beta (\alpha_i - \alpha_j)} A_{(i,j)} = U_{\beta} A_{(i,j)} U_{\beta}^{\dagger}
\end{equation*}
so the MPS generated by means of the Kraus operators $\mathcal{K}$
has the local symmetry $G$. Moreover, the MPS is trivially injective
when $D \ge 3$. We can prove this by choosing arbitrary $k$ and
$k'$. Since $D \ge 3$, we can always find an $l$ such that $k' \neq
l \neq k$ and then $|k \rangle \langle k' | = |k \rangle \langle l |
l \rangle \langle k' | = A_{(k,l)} A_{(l, k')}$.
\end{ex}

Let us remark that this counter-example is applicable to spin $\ge
\frac{5}{2}$. Indeed, one can prove Theorem \ref{} for $U(1)$ and
spin $\frac{1}{2}$, which is the content of the following
proposition. The case of spin $3/2$ remains an open question.
\begin{prop}
If $|\Phi \rangle$ is an MPS with physical dimension $d=2$ and
invariant under $U(1)$, then $|\Phi \rangle$ cannot be injective.
\end{prop}
\begin{proof}
We will show it by contradiction. By choosing a basis where the
physical unitary $u$ is diagonal, the condition on the Kraus
operators becomes
\begin{equation*}
e^{i \lambda_n \phi} A_n = e^{i H \phi} A_n e^{-i H \phi}
\end{equation*}
where $H$ is the hermitian generator of the symmetry. Let us expand
the expression for infinitesimal angles
\begin{equation*}
[H,A_n] = \lambda_n A_n
\end{equation*}
which is the equation of eigenvalues for the operator $L(\bullet) =
[H,\bullet]$. This can be transformed into an ordinary
eigenvalue equation for the matrix operator
$L = H \otimes \1 - \1 \otimes \bar{H}$. The diagonalization can be easily
performed by taking the spectral decomposition of $H = \sum_i \mu_i
P_i$, where $P_i$ are orthogonal projectors. It straightforwardly
follows that the eigenvalues of $L$ are $\lambda_{i j} = \mu_i -
\mu_j$ and the corresponding eigenoperators fulfil $A_{i j} = P_i
A_{i j} P_j$.

Let us focus now on the case $d=2$. Then, we have that $A_1 = P_1
A_1 P_{\alpha}$ and $A_2 = P_{\beta} A_2 P_{\gamma}$ for some
$\alpha, \beta, \gamma$. If $\beta=1$  $P_1X=X$ for all $X\in {\rm
span}\{A_{i_1}\cdots A_{i_n}\}$ and the MPS cannot be injective. The
same happens if $\alpha=\gamma$. So let us assume that $\beta\not =
1$ and $\gamma \not = \alpha$. Now if $\alpha=1$, we have
$A_1=P_1A_1P_1$, $A_2=(\1-P_1)A_2(\1-P_1)$ and the MPS is block
diagonal and hence non-injective. The same happens if
$\beta=\gamma$. So $\alpha\not =1$ and $\beta\not =\gamma$ and this
gives $A_1^2=0=A_2^2$ which implies that ${\rm span}\{A_{i_1}\cdots
A_{i_n}\}={\rm span}\{A_1A_2A_1A_2\cdots, A_2A_1A_2A_1\cdots\}$ has
dimension $\le 2$.
\end{proof}

\end{section}


\begin{section}{General construction of $SU(2)$
two-body Hamiltonians with MPS eigenstates}\label{Sec:TBH}

We have seen in Definition \ref{parham} a way, called the parent
Hamiltonian method, to construct local $SU(2)$-symmetric
Hamiltonians with MPS as eigenstates. In this section we  first
prove that this method is the most general one to find Hamiltonians
having a given MPS as {\it local eigenstate}, that is, being an
eigenstate of each local term in the Hamiltonian. Then, we  show
examples (including the AKLT and Majumdar-Ghosh states) of MPS that
are {\it excited} eigenstates of local two-body translationally
invariant $SU(2)$-symmetric Hamiltonians. More examples are then
provided in the appendix.


\begin{subsection}{Completeness of the parent Hamiltonian method}

\begin{thm}
Given an MPS $|\psi\>$, any translational invariant Hamiltonian
having it as a {\emph local eigenstate} is of the form $a\1 + H$
where $H$ is a parent Hamiltonian for $|\psi\>$ in the sense of
Definition \ref{parham}.
\end{thm}

\begin{proof}
Let us call $h$ the local hamiltonian. By hypothesis of local
eigenstate,
\begin{equation}\label{descham}
h\rho=\lambda \rho
\end{equation}
for certain $\lambda \in \R$. This implies  $[\rho,h]=0$ and hence
one can find a set of projectors $\mathcal{P}= \{ P_i, i=1,\ldots, r
| \sum_i P_i = \1 \}$ such that we can decompose both $\rho$ and $h$
by means of them, i.e. $h = \sum_{i} a_i P_i$ and $\rho = \sum_{j
\in C} b_j P_j$, where $C$ represents the set of projectors which
describe the support of $\rho$. Using eq. (\ref{descham}) with this
decomposition gives that $a_i=\lambda$ for all $i\in C$ and hence
\begin{equation*}
\begin{split}
h & = \sum_{i \in C^{\perp}} a_i P_i + \lambda \sum_{i \in C} P_i \\
& = \sum_{i \in C^{\perp}} (a_i - \lambda) P_i + \lambda \1\;.
\end{split}
\end{equation*}
Then, the translational invariance hamiltonian is $H=\sum_j
\tau^j(h)\otimes \1_{\rm rest}$, where $\tau$ is the translation
operator. The theorem follows from replacing the result for the
local hamiltonian and comparing this with Definition \ref{parham} of
parent hamiltonian.
\end{proof}

This Theorem shows that, given an MPS $|\psi\>$, looking for all
possible parent Hamiltonians of interaction length $k$ is equivalent
to look for all possible solutions to the equation
\begin{equation}\label{eq:eigen-local}h\rho^{(k)}=\lambda
\rho^{(k)},\end{equation} with $\lambda =\tr{h\rho^{(k)}}$. The next
lemma gives yet another equivalent formulation, which is the one we
will use in the sequel.

\begin{lem}
Given a Hermitian matrix $h$ and a density matrix $\rho$,
$h\rho=\lambda \rho$ if and only if \begin{equation}\label{eigcond}
\tr{h^{2} \rho} - \tr{h \rho}^2 = 0\;.
\end{equation}
\end{lem}
\begin{proof}
One implication is clear. For the other, let us write $\<h\>$ for
$\tr{h\rho}\1$. By assumption
$$\tr{(h-\<h\>)^2\rho}=\tr{h^2\rho}-\tr{h\rho}^2=0.$$ So
$\rho^{1/2}(h-\<h\>)^2\rho^{1/2}=0$, since it is a positive operator
with trace $0$. This implies that $(h-\<h\>)\rho=0$ and hence
$h\rho=\lambda \rho$.
\end{proof}

With this at hand we can systematically search for MPS that are
excited {\it local eigenstates} of $SU(2)$ invariant Hamiltonians
with two-body interactions. We will proceed as follows. We start
with a given $SU(2)$ symmetric MPS $|\psi\>$  and fix the
interaction length $n$. Then we look for possible solutions to Eq.
(\ref{eigcond}) of the form \begin{equation}\label{SU2genham} h =
\sum_{i<j\leq n} \sum_{\alpha = 1}^{2 J} a_{i j}^{(\alpha)}
(\vec{S}_i \circ \vec{S}_j)^{\alpha} + a_0 \1\;,
\end{equation}
to ensure $SU(2)$ symmetry and two body interactions in the
Hamiltonian. Finally, to guarantee that the MPS $|\psi\>$ is an {\it
excited} state, we will find another $SU(2)$ symmetric MPS with less
energy that will act as a witness. In the next section we will
illustrate this procedure starting with $|\psi\>$ the AKLT, the
Majumdar-Ghosh state, and generalizations. Throughout we work in the
thermodynamical limit $N\rightarrow \infty$.
\end{subsection}


\begin{subsection}{Examples of $SU(2)$ two-body Hamiltonians}

\begin{subsubsection}{Spin 1}
Let us consider the AKLT state as a first example. Its Kraus
operators are $A_{-1}=-\sqrt{2}\sigma^- $, $A_0=\sigma^z$,
$A_1=\sqrt{2}\sigma^+$.

In the case $n=2$ the only solution to Eq. (\ref{eigcond}) is the
AKLT Hamiltonian. In the case  $n = 3$, the solutions are given by
\begin{multline*}
h = (-3 v_1 + v_2 +3 v_3)(\vec{S}_1 \circ \vec{S}_2)+ v_3 (\vec{S}_1
\circ \vec{S}_2)^2 + \\ \frac{1}{2}(-3 v_1 + v_2)(\vec{S}_1 \circ \vec{S}_3)
-  \frac{1}{2}(-3 v_1 + v_2)(\vec{S}_1 \circ \vec{S}_3)^2 + \\ v_2
(\vec{S}_2 \circ \vec{S}_3) + v_1 (\vec{S}_2 \circ \vec{S}_3)^2
\end{multline*}
where the eigenvalue corresponding to the AKLT state is $7 v_1 - 3
v_2 - 2 v_3$. The total translational invariant Hamiltonian is then
\begin{multline*}
H = \sum_i (-3 v_1 + 2 v_2 +3 v_3)(\vec{S}_i \circ \vec{S}_{i + 1}) + \\
(v_1 + v_3) (\vec{S}_i \circ \vec{S}_{i + 1})^2 + \frac{1}{2}(-3 v_1 + v_2)
(\vec{S}_i \circ \vec{S}_{i+2}) - \\ \frac{1}{2}(-3 v_1 + v_2)
(\vec{S}_i \circ \vec{S}_{i+2})^2
\end{multline*}
which contains the usual AKLT model. It is not difficult to check
that there is a region in the parameter space where the AKLT state
is still the ground state of this Hamiltonian. To find regions where
it is an {\it excited}  eigenstate we will use as a witness the
$SU(2)$ symmetric MPS associated to the virtual representation
$\frac{3}{2}\oplus\frac{1}{2}$ (see Section \ref{2}). The result is
plotted in fig. \ref{AKLT}, where one sees the existence of points
in this family of spin $1$ Hamiltonians for which the AKLT state is
an excited state.

\begin{figure}[h!]
 \begin{center}
  \includegraphics[width=0.4\textwidth]{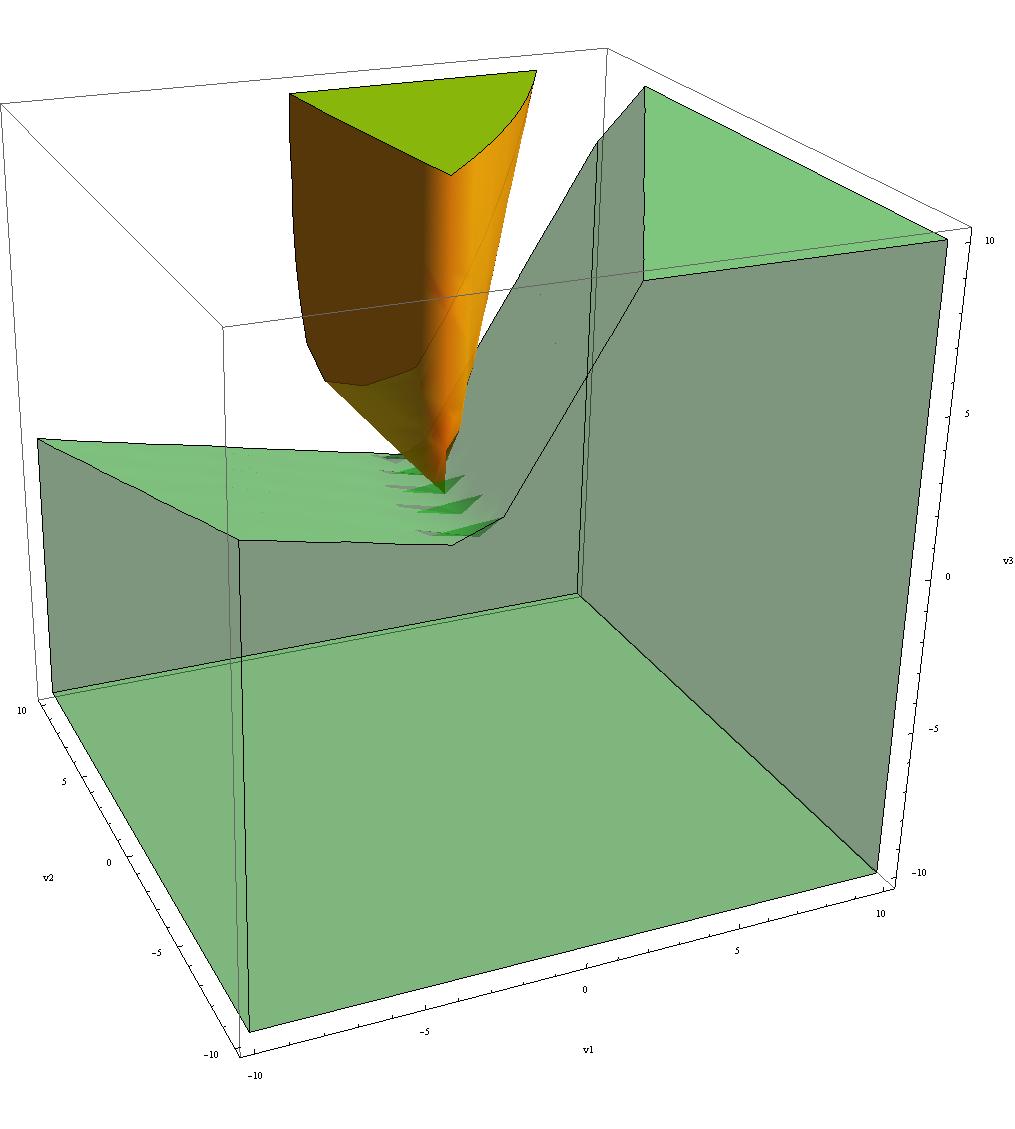}
  \caption[Fig. 3]{\footnotesize \emph{Space of parameters of the local Hamiltonian $h$ for the
   AKLT state and $n=3$. The orange volume represents the points where the
   state is the local (and hence the global) ground state. The green volume
   represents points corresponding to excited states detected with
   the witness $\frac{3}{2} \oplus \frac{1}{2}$.}}
  \label{AKLT}
 \end{center}
\end{figure}

Note that it is possible to perform a change of variables in the
total Hamiltonian, for instance $a\rightarrow\frac{1}{2}(-3v_1 +
v_2)$ and $b\rightarrow v_1 + v_3$, such that it depends only on two
parameters. However, the number of parameters that the local
Hamiltonian $h$ depends on cannot be reduced, which means that there
are non-physical parameters in it. In Fig. \ref{AKLTphys} we have
represented the problem above ($n=3$ and AKLT state) in terms of the
physical parameters. The positive axis $b$ corresponds there to the
usual AKLT Hamiltonian.

\begin{figure}[h!]
 \begin{center}
  \includegraphics[width=0.4\textwidth]{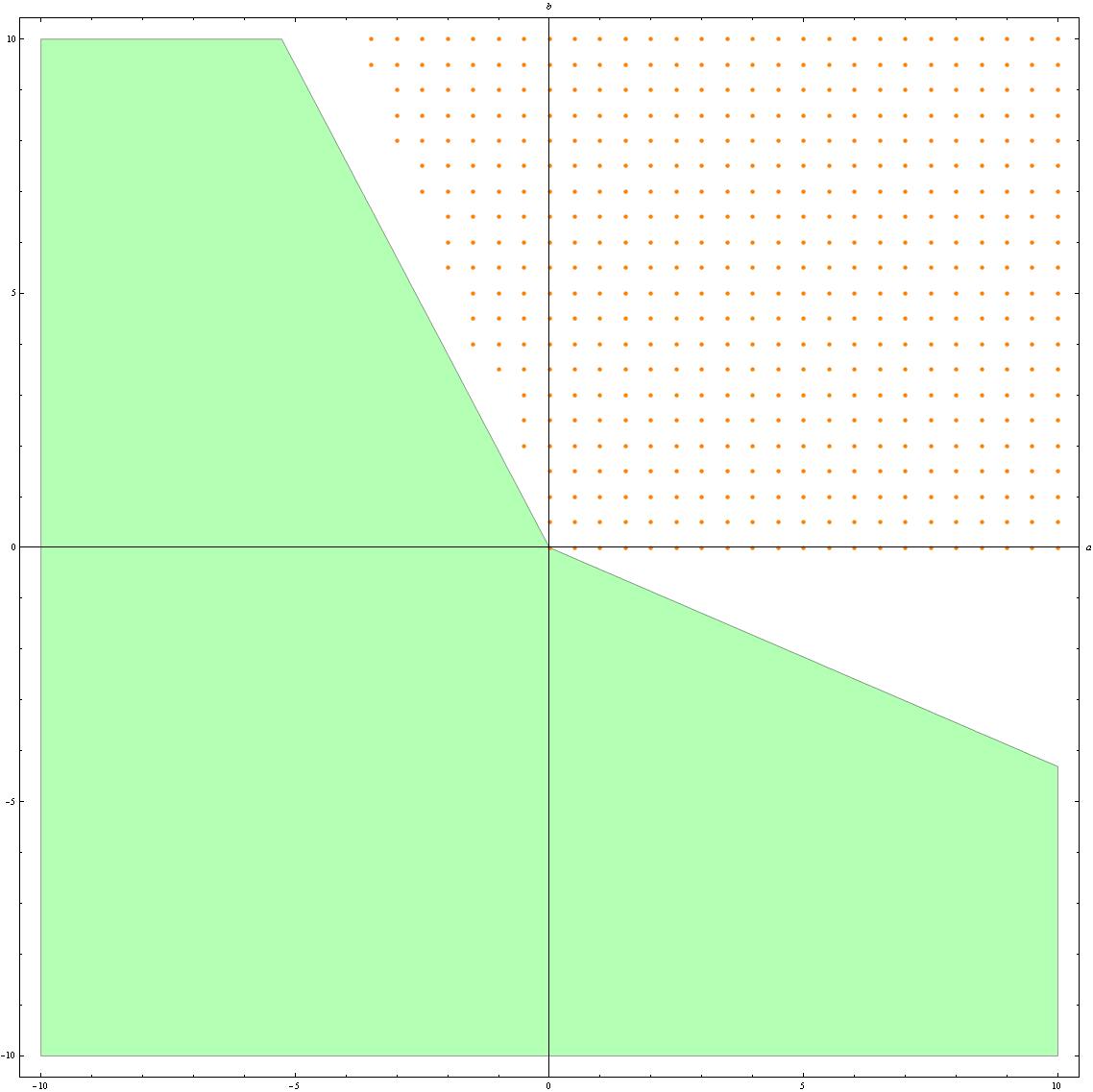}
  \caption[Fig. 4]{\footnotesize \emph{Space of physical parameters
  of the global Hamiltonian $H$ corresponding to $n=3$ and the AKLT state. The orange points represent where the state is
  the local (and hence the global) GS. The green surface represents
  points corresponding to excited states detected by means of the
  witness $\frac{3}{2} \oplus \frac{1}{2}$.}}
  \label{AKLTphys}
 \end{center}
\end{figure}

Concerning FNW states, that is integer spin $J$ and virtual irrep
$j$, we have  performed an exhaustive search and table \ref{table1}
gathers the main results. The study has been carried out by
increasing $n$ and studying the number of parameters which the
family of Hamiltonians depends on (notice that the case of
interaction length $n$ contains the case of interaction length
$n-1$). We have increased $n$ until the number of parameters stops
growing. In all the cases considered in the table, a saturation
occurs when $n > 3$, i.e. considering more than $3$
particles does apparently not add new Hamiltonians. \\

\begin{table}
\begin{center}
\begin{tabular}{|c|c|c|c|c|c|c|c|}
\hline
\backslashbox{\phantom{a} J}{j \phantom{a}} & $\phantom{a} \frac{1}{2} \phantom{a}$ & $\phantom{a} 1 \phantom{a}$ & $\phantom{a} \frac{3}{2} \phantom{a}$ & $\phantom{a} 2 \phantom{a}$ & $\phantom{a} \frac{5}{2} \phantom{a}$ & $\phantom{a} 3 \phantom{a}$ & $\phantom{a} \frac{7}{2} \phantom{a}$\\
\hline
1 & 2 & 1 & $\blacksquare$ & $\blacksquare$ & $\blacksquare$ & $\blacksquare$ & $\blacksquare$ \\
\hline
2 & --- & 5 & 3 & $\blacksquare$ & $\blacksquare$ & $\blacksquare$ & $\blacksquare$ \\
\hline
3 & --- & --- & 4 & 2 & 2 & 1 & $\blacksquare$ \\
\hline
\end{tabular}
\end{center}
\caption{\footnotesize \emph{Table of results for FNW states with physical spin $J$ and virtual spin $j$. The numbers in the table are
 the number of parameters the obtained families of Hamiltonians
 depend on. The $\blacksquare$ represent the cases for which
 no solution was found.}} \label{table1}
\end{table}

Let us also introduce a new state of spin $1$ with virtual spin $1$, given by the Kraus operators
$$A_{1}= \frac{1}{\sqrt{2}}\left(
                     \begin{array}{ccc}
                       0 & 1 & 0 \\
                       0 & 0 & 1 \\
                       0 & 0 & 0 \\
                     \end{array}
                   \right)
\quad ,\quad A_{0}= \frac{1}{\sqrt{2}}\left(
                     \begin{array}{ccc}
                       1 & 0 & 0 \\
                       0 & 0 & 0 \\
                       0 & 0 & -1 \\
                     \end{array}
                   \right)$$
$$A_{-1}= \frac{1}{\sqrt{2}} \left(
                     \begin{array}{ccc}
                       0 & 0 & 0 \\
                       -1 & 0 & 0 \\
                       0 & -1 & 0 \\
                     \end{array}
                   \right)$$
The total translational invariant hamiltonian which has this state
as eigenstate is
\begin{multline*}
H = \sum_i (\vec{S}_i \circ \vec{S}_{i + 1})^2 -
(\vec{S}_i \circ \vec{S}_{i+2}) - (\vec{S}_i \circ \vec{S}_{i+2})^2
\end{multline*}
This state is injective and a local excited
state. The fact that this state is an excited state of the global
hamiltonian can be checked as above by means of the witness $1
\oplus 0$.
\end{subsubsection}
\newpage
\begin{subsubsection}{Spin $\frac{1}{2}$}
Let us consider now the the Majumdar-Ghosh state as an example with
semi-integer spin. The Kraus operators are now
$$A_{-\frac{1}{2}}=\left(
                     \begin{array}{ccc}
                       0 & \frac{1}{\sqrt{2}} & 0 \\
                       0 & 0 & -1 \\
                       0 & 0 & 0 \\
                     \end{array}
                   \right)
 \quad ,\quad A_{\frac{1}{2}}=\left(
                     \begin{array}{ccc}
                       0 & 0 & 0 \\
                       1 & 0 & 0 \\
                       0 & \frac{1}{\sqrt{2}} & 0 \\
                     \end{array}
                   \right)$$
As in the previous case, we do not find any solution for $n = 2$ and
only the Majumdar-Ghosh  Hamiltonian for the cases $n = 3$ and $n =
4$. For $n = 5$ the solutions to Eq. (\ref{eigcond}) are given by
\begin{multline}\label{blockH2}
h = \tiny{(v_1-v_2+v_4)}(\vec{S}_1 \circ \vec{S}_2)+ (v_1-v_2+v_4)
 (\vec{S}_1 \circ \vec{S}_3) + \\ v_3(\vec{S}_1 \circ \vec{S}_4) +
 v_3(\vec{S}_1 \circ \vec{S}_5) + v_4 (\vec{S}_2 \circ \vec{S}_3) \\
 + (-v_1+v_2+v_3) (\vec{S}_2 \circ \vec{S}_4)+ v_3(\vec{S}_2 \circ
  \vec{S}_5) + v_2(\vec{S}_3 \circ \vec{S}_4)+ \\ v_1(\vec{S}_3
  \circ \vec{S}_5) + v_1(\vec{S}_4 \circ \vec{S}_5)
\end{multline}
and the energy associated to the state is $-\frac{3}{4}(v_1 + v_4)$.
The total Hamiltonian $H = \sum_i \tau_i(h)$ is given by
\begin{multline}
H = \sum_i 2(v_1+v_4)(\vec{S}_i \circ \vec{S}_{i + 1}) + (v_1 +
v_3 + v_4) (\vec{S}_i \circ \vec{S}_{i + 2}) \\ + 2v_3(\vec{S}_i
\circ \vec{S}_{i+3}) + v_3(\vec{S}_i \circ \vec{S}_{i+4})
\end{multline}

As in the AKLT case, by means of a change of variables $a
\rightarrow v_3$ and $b \rightarrow v_1 + v_4$, the number of
physical parameters in the total Hamiltonian is $2$, compared with
the four parameters the local Hamiltonian depends on. The
Majumdar-Ghosh state is an excited local eigenstate for a region in
the space of parameters, which in this case is detected by the
witness $\frac{1}{2} \oplus 1 \oplus 0$, as shown in fig.
$\ref{MG}$. The usual Majumdar-Ghosh Hamiltonian \footnote{The family
of hamiltonians constructed in \cite{Ku02} is quite remarkable. In this
paper the ground state of the hamiltonian which corresponds to $b=2a$
is calculated, and fits perfectly to our results.} corresponds to
the positive axis $b$.

\begin{figure}[h!]
 \begin{center}
  \includegraphics[width=0.4\textwidth]{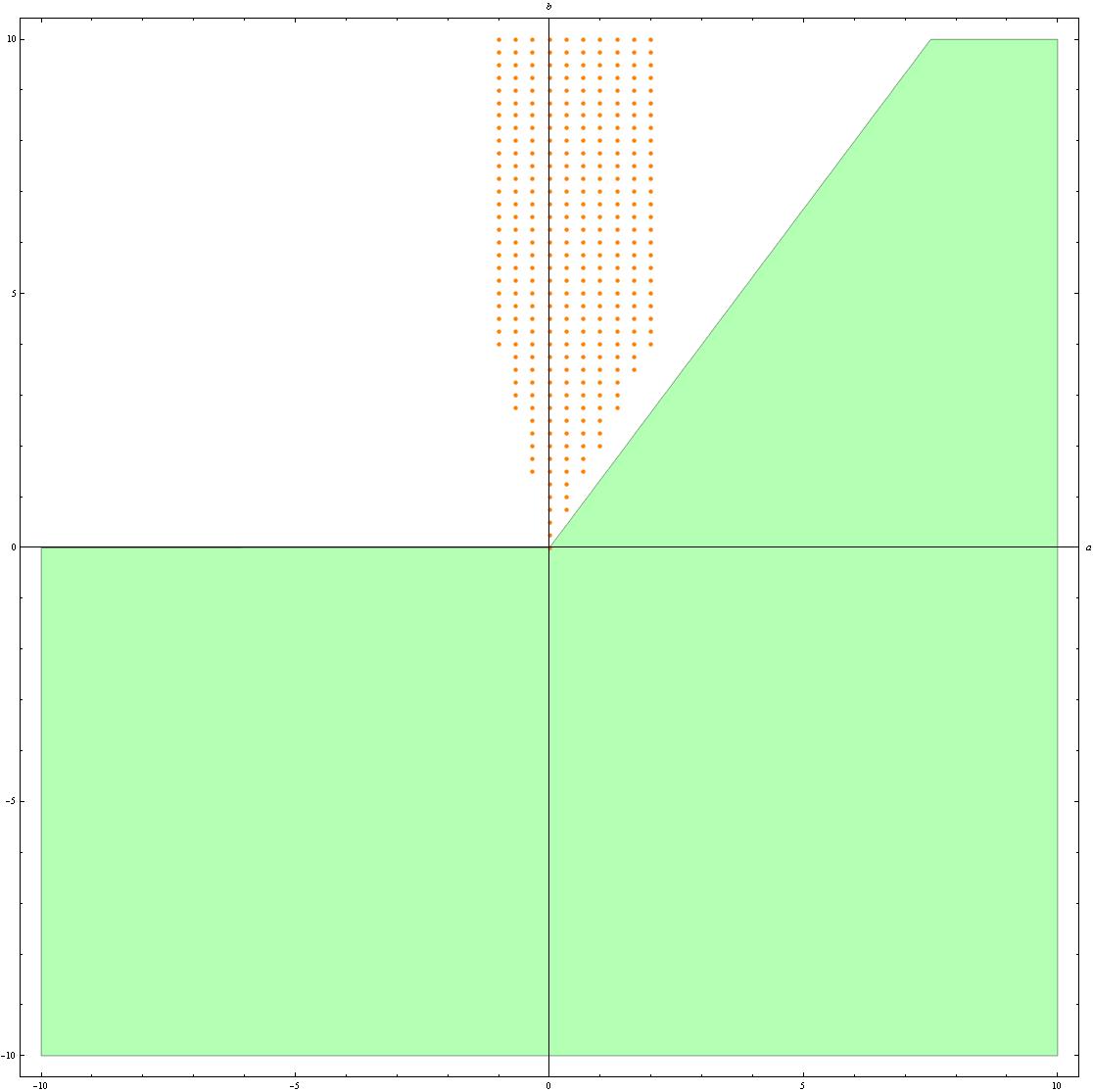}
  \caption[Fig. 5]{\footnotesize \emph{Space of physical parameters
  of the total Hamiltonian for $n=5$ associated to the Majumdar-Ghosh state. The orange points represent where
  the state is the local (and hence the global) ground state. The green
  surface represents points corresponding to excited states
  detected by means of the witness $\frac{1}{2} \oplus 1 \oplus 0$.}}
  \label{MG}
 \end{center}
\end{figure}
\end{subsubsection}

\begin{subsubsection}{Spin $\frac{3}{2}$}
Let us consider as final example the $SU(2)$ symmetric MPS
corresponding to spin $\frac{3}{2}$ and virtual representation
$\frac{3}{2} \oplus 0$. For $n=3$, the solutions to Eq.
(\ref{eigcond}) are given by
\begin{multline}\label{blockH2}
h = v_3(\vec{S}_1 \circ \vec{S}_2)+ v_2 (\vec{S}_1 \circ \vec{S}_2)^2
+ v_1(\vec{S}_1 \circ \vec{S}_2)^3 + \\ (2v_1 -v_2 + v_3)(\vec{S}_1
 \circ \vec{S}_3) + (4v_1 - v_2) (\vec{S}_1 \circ \vec{S}_3)^2 \\ +
 v_1(\vec{S}_1 \circ \vec{S}_3)^3 + v_3(\vec{S}_2 \circ \vec{S}_3) +
  v_2(\vec{S}_2 \circ \vec{S}_3)^2 + \\ v_1(\vec{S}_2 \circ \vec{S}_3)^3
\end{multline}
and the energy associated to the MPS is in this case
$-\frac{15}{64}(165 v_1 - 60 v_2 + 16 v_3)$. The global Hamiltonian
reads now
\begin{multline}
H = \sum_i 2v_3(\vec{S}_i \circ \vec{S}_{i + 1}) + 2v_2 (\vec{S}_i
\circ \vec{S}_{i + 1})^2 \\ + 2v_1(\vec{S}_i \circ \vec{S}_{i+1})^3
+ (2v_1 -v_2+v_3)(\vec{S}_i \circ \vec{S}_{i+2})+ \\ (4v_1 -v_2)
(\vec{S}_i \circ \vec{S}_{i+2})^2+v_1(\vec{S}_i \circ \vec{S}_{i+2})^3
\end{multline}
It is remarkable that in this case there are no spurious parameters
in the local Hamiltonian $h$. Considering the family of states whose
virtual representation is $\frac{3}{2} \oplus 1 \oplus 0$ as a
witness, it is possible to demonstrate that there is a region in the
space of parameters of the Hamiltonian for which the MPS is an
excited eigenstate, as shown in Fig.~\ref{32}.

\begin{figure}
  \includegraphics[width=0.4\textwidth]{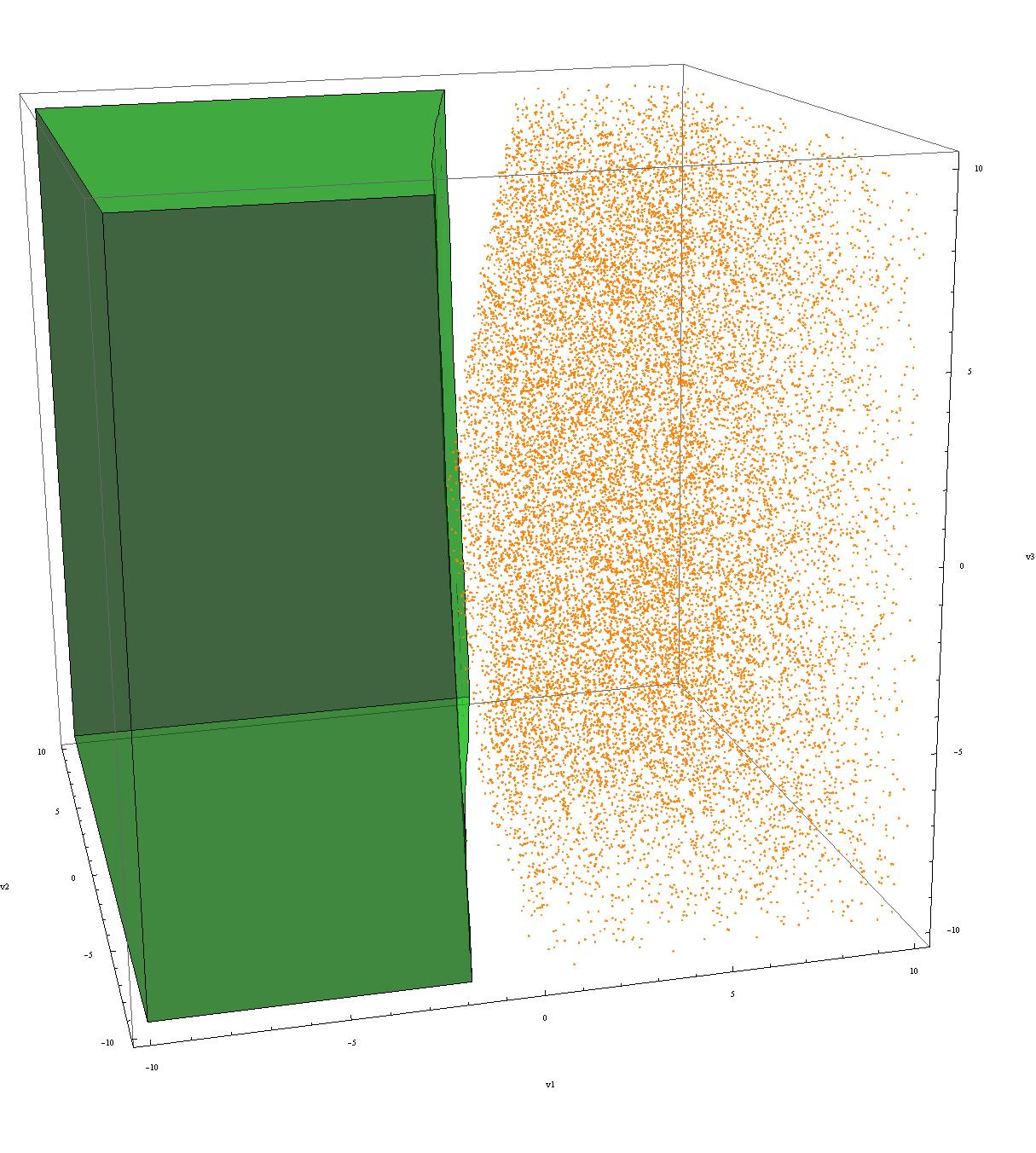}
  \caption{\footnotesize \emph{Space of parameters of the
  spin $\frac{3}{2}$ model. The orange points are obtained numerically
  and they represent values of the parameters where the MPS state
  is the GS. The green volume represents points corresponding to
  excited states detected with the witness $\frac{3}{2} \oplus 1 \oplus 0$.}}
  \label{32}
\end{figure}
\end{subsubsection}
\end{subsection}
\end{section}


\begin{section}{Conclusions}
Despite the fact that all our results are restricted to the family
of TIMPS, their relevance is manifested by the fact that those
states approximate all ground states of 1-dimensional Hamiltonians
with short range interactions. Thus, one would expect that the
properties derived for MPS would be relevant in a more general
context. Moreover, due to their simplicity, MPS can be then thought
as a 'laboratory' where to search for some generic mathematical and
physical properties of states that are relevant in 1-dimensional
spin chains. Later on, one may use more powerful mathematical
methods to try to extrapolate those properties to general spin
chains. Furthermore, many of the techniques used in the present work
are amenable of an extension to higher spatial dimensions, where
PEPS play the role of MPS. In Ref. \cite{PWSVC08} some first results
in this direction were derived, which will be generalized in a
further publication.
\end{section}


\begin{section}{Acknowledgments}
M. Sanz would like to thank Dr. Miguel Aguado, Dr. Marco Roncaglia
and Dr. Frank Verstraete for their helpful discussions and the QCCC
Program of the EliteNetzWerk Bayern as well as the DFG (FOR 635, MAP
and NIM) for the support. D. Perez-Garcia acknowledges financial
support from Spanish grants I-MATH, MTM2005-00082 and
CCG07-UCM/ESP-2797 and M.M. Wolf acknowledges support by QUANTOP and
the Danish Natural Science Research Council(FNU).
\end{section}


\appendix


\begin{section}{Relations between irreducibility and injectivity}

In this appendix we  give a direct proof of the fact that an
irreducible representation in the virtual level of a symmetric MPS
implies that the MPS is injective. We  also see that the reverse
inclusion is not true in general, but it holds under some conditions
on the Kraus operators.

We have to recall that, given a set of Kraus operators defining an
MPS $\mathcal{K}=\{A_1,\ldots, A_d\}$,  we can define an associated
{\it completely positive map}
$\mathbb{E}(X)=\sum_{i=1}^dA_iXA_i^\dagger$. The symmetry in the MPS
transfers then to the {\it covariance} of the channel, that is,
$\mathbb{E}(U_gXU_g^\dagger)=U_g\mathbb{E}(X)U_g^\dagger$ for all
$X$. It is shown in \cite{FNW92, PVWC07} that if $\mathbb{E}$ is
trace preserving and has $\1$ as its  unique fixed point, then the
MPS is injective. Moreover, it is trivial to see that if
$\mathbb{E}$ is the ideal channel ($\mathbb{E}(X)=X$ for all X),
then the MPS is a product state. Therefore, the desired result that
{\it irrep implies injectivity} is a consequence of the following
theorem.

\begin{thm}\label{irrepinj}
Let us take a completely positive map $\E : \mathcal{M}_{D} \lra
\mathcal{M}_{D}$ that is covariant for an irrep of a compact
connected Lie group $G$. Then, either $\E$ is the ideal channel or
it is trace preserving and the identity its unique fixed point.
\end{thm}

\begin{proof}
Let us consider a fixed point $\Delta$ of $\E$. Then $U_g \Delta
U_g^{\dagger}$ is also a fixed point because of the covariance.
Therefore, integrating under the Haar measure and using Schur's
lemma, $\1$ is also a fixed point. A similar argument shows that
$\E$ is also trace preserving.

Now we can apply L\"uders' theorem \cite{BJKW00}, which  ensures
that the set of fixed points $\mathcal{P}$ of  $\E$ coincides with
the commutant $\mathcal{K}'$ of the set of Kraus operators of $\E$.
This is trivially a $C^*$-subalgebra of $\mathcal{M}_{D}$. Moreover,
we know by the classification of the $C^*$-subalgebras in
$\mathcal{M}_{D}$ that there exists a unitary $V\in \mathcal{M}_D$
such that $V \mathcal{P} V^{\dagger} = \oplus_i (M_{n_i} \otimes
\1_{n'_i})=\mathcal{A}$.

The equivalent representation $V_g = V U_g V^{\dagger}$ is also an
irrep and fulfils that $V_g \mathcal{A} V_g^{\dagger} =
\mathcal{A}$. This means that the block structure of $\mathcal{A}$
remains invariant under the action of $V_g$ by conjugation. Now we
use that
\begin{equation}\label{eq4}
V_g \mathcal{A} V_g^{\dagger} \subset \mathcal{A} \Leftrightarrow [J,\mathcal{A}]\subset \mathcal{A} \text{ for all generators } J.
\end{equation}
This implies that $J$ has the same block structure as $\mathcal{A}$.
If there is more than one block, the representation is reducible. If
$\mathcal{A} = M_{n}\otimes \1_{n'}$, then we use again eq.
$\eqref{eq4}$:

The Schmidt decomposition allows us to take $J = \sum_{i} A_i
\otimes B_i$ where the $B_i$'s form a basis of $M_{n'}$, with
$B_1=\1$. Then, eq. \ref{eq4} gives that $\sum_i [A_i,M_{n}]\otimes
B_i=C\otimes \1$, which implies that $A_i$ is proportional to $\1$
for all $i\ge 2$. This gives $J=\1\otimes X + Y\otimes \1$ and hence
$V_g=V_1^g\otimes V_2^g$, which is reducible unless $\mathcal{A}=\1$
or $\mathcal{A}=M_N$ (which implies that $\E$ is the ideal channel).
\end{proof}

Although the implication in the opposite direction could also seem
true, it is not, as shown by the following example.
\begin{ex}
Let us consider the family of $SU(2)$ symmetric MPS of spin $1$ with a reducible
virtual representation $\frac{1}{2} \oplus \frac{3}{2}$ given by the following maps (see Section \ref{2}).
\begin{equation*}
  \tilde{V} =
        \begin{pmatrix}
           e^{i \alpha_{1 1}} \cos{\theta_1} V_{\frac{1}{2}}^{\frac{1}{2}} & e^{i \alpha_{1 2}} \sin{\theta_2} V_{\frac{1}   {2}}^{\frac{3}{2}} \\
           \phantom{u} & \phantom{u} \\
           e^{i \alpha_{2 1}} \sin{\theta_1} V_{\frac{3}{2}}^{\frac{1}{2}} & e^{i \alpha_{2 2}} \cos{\theta_2} V_{\frac{3}{2}}^{\frac{3}{2}} \\
        \end{pmatrix}
\end{equation*}
It is not difficult to check that the MPS is injective except in
particular directions in space, such as those for which the isometry
breaks into blocks, i.e. $\theta_i = n \frac{\pi}{2}$.
\end{ex}

Although the equivalence is not true in general, we can still give a
sufficient condition which applies, for instance, to the AKLT and
other FNW states. Let us recall from \cite{PWSVC08} or Theorems
\ref{thm:MPS-sym} and \ref{thm:G-sym} that an injective symmetric
MPS verifies
\begin{equation}\label{eq:inj-ir}
\sum_{i} u_{ij}^gA_i=e^{i\theta_g}U_gA_jU_g^\dagger\; ,
\end{equation}
where in addition one may ask for $\sum_{i}A_i^\dagger A_i=\1$ \cite{PVWC07}.

\begin{prop}\label{injimpirrep}
If $u_g$ is irreducible and $\{A_i^\dagger A_j\}_{i,j}$ spans the
whole space of matrices, then the virtual representation $U_g$ of
(\ref{eq:inj-ir}) is also irreducible.
\end{prop}
\begin{proof}
From (\ref{eq:inj-ir}) one gets
\begin{equation*}
\sum_{i_1 , i_2} \bar{u}_{i_1 j_1}^g u_{i_2 j_2}^g A_{i_1}^{\dagger} A_{i_2} = U_g A_{j_1}^{\dagger}A_{j_2} U_g^{\dagger}\; .
\end{equation*}
Integrating now with respect to the Haar measure, the lhs is
simplified by the irreducibility of $u_g$ and the orthogonality
relations. The result is $\delta_{j_1 j_2}\sum_i A_i^{\dagger} A_i =
\delta_{j_1 j_2} \1$. This means that $\int_G U_g X U_g^{\dagger}
\propto \1$, $\forall X \in \mathcal{M}_D$, since we can span the
complete space of matrices. But this implies that $U_g$ is an irrep
by means of the inverse of Schur's lemma.
\end{proof}
\end{section}

\begin{section}{List of parent Hamiltonians}
The following lists $SU(2)$-invariant two-body Hamiltonians for
which the MPS with physical spin $J$ (irrep) and virtual spin $j$ is
an exact eigenstate with energy $\epsilon$.
\begin{subsection}{Spin $J =
\frac{1}{2}$}
$\bullet \phantom{A} j = \frac{1}{2} \oplus 0$, $\epsilon = -\frac{3}{4} (v_1 + v_4)$:\\
\begin{multline*}
H = \sum_i 2(v_1+v_4)(\vec{S}_i \circ \vec{S}_{i + 1}) + (v_1 + v_3 + v_4) (\vec{S}_i \circ \vec{S}_{i + 2}) \\ + 2v_3(\vec{S}_i \circ \vec{S}_{i+3}) + v_3(\vec{S}_i \circ \vec{S}_{i+4})
\end{multline*}

$\bullet$ No solutions found (with $n\leq 6$) for $ j = \frac{1}{2} \oplus 1$, $\frac{3}{2} \oplus 1, \frac{3}{2} \oplus 2, \frac{5}{2} \oplus 2.$\\
\end{subsection}

\begin{subsection}{Spin $J = 1$}
$\bullet \phantom{A} j = \frac{1}{2}$, $\epsilon = 7 v_1 - 3 v_2 -2 v_3$:\\
\begin{multline*}
H = \sum_i (-3 v_1 + 2 v_2 +3 v_3)(\vec{S}_i \circ \vec{S}_{i + 1}) + \\ (v_1 + v_3) (\vec{S}_i \circ \vec{S}_{i + 1})^2 + \frac{1}{2}(-3 v_1 + v_2)(\vec{S}_i \circ \vec{S}_{i+2}) - \\ \frac{1}{2}(-3 v_1 + v_2)(\vec{S}_i \circ \vec{S}_{i+2})^2
\end{multline*}

$\bullet \phantom{A} j = 1$, $\epsilon = 1$:\\
\begin{multline*}
H = \sum_i (\vec{S}_i \circ \vec{S}_{i + 1})^2 -(\vec{S}_i \circ \vec{S}_{i+2}) -(\vec{S}_i \circ \vec{S}_{i+2})^2
\end{multline*}

$\bullet \phantom{A}$ No solutions found (with $n\leq4$) for $ j = \frac{3}{2},2
, \frac{5}{2}, 3$.\\

\end{subsection}

\begin{subsection}{Spin $J = \frac{3}{2}$}
$\bullet \phantom{A} j = \frac{3}{2} \oplus 0$, $\epsilon = -\frac{15}{64}(165 v_1 - 60 v_2 +16 v_3)$:\\
\begin{multline*}
H = \sum_i 2 v_3(\vec{S}_i \circ \vec{S}_{i + 1}) + 2 v_2 (\vec{S}_i \circ \vec{S}_{i + 1})^2 + \\ 2 v_1 (\vec{S}_i \circ \vec{S}_{i + 1})^3 + (2 v_1 - v_2 + v_3)(\vec{S}_i \circ \vec{S}_{i+2}) + \\ (4 v_1 - v_2)(\vec{S}_i \circ \vec{S}_{i+2})^2 + v_1 (\vec{S}_i \circ \vec{S}_{i+2})^3
\end{multline*}

$\bullet \phantom{A} j = \frac{1}{2} \oplus 1$, $\epsilon = -\frac{495}{64}$:\\
\begin{multline*}
H = \sum_i \frac{243}{16}(\vec{S}_i \circ \vec{S}_{i + 1}) + \frac{29}{4} (\vec{S}_i \circ \vec{S}_{i + 1})^2 + \\  (\vec{S}_i \circ \vec{S}_{i + 1})^3
\end{multline*}

$\bullet$ No solutions found (with $n\leq 4$) for $j = \frac{3}{2} \oplus 1$,$ \frac{5}{2} \oplus 1, \frac{1}{2} \oplus 2, \frac{3}{2} \oplus 2$.
\end{subsection}

\begin{subsection}{Spin $J = 2$}
$\bullet \phantom{A} j = 1$, $\epsilon = (-6986 v_1 + 778 v_2 - 62 v_3 + 1260 v_4 - 90 v_5)$:\\
\begin{multline*}
H = \sum_i (2400 v_1 - 63 v_2 + 24 v_3 -792 v_4 + 63 v_5)(\vec{S}_i \circ \vec{S}_{i + 1}) + \\ (133 v_1 - 14 v_2 + 2 v_3 - 133 v_4 + 14 v_5) (\vec{S}_i \circ \vec{S}_{i + 1})^2 + \\ (v_2 + v_5) (\vec{S}_i \circ \vec{S}_{i + 1})^3 + (v_1 + v_4)(\vec{S}_i \circ \vec{S}_{i + 1})^4 + \\ (\frac{1729}{2} v_1 - 91 v_2 + \frac{13}{2} v_3) (\vec{S}_i \circ \vec{S}_{i+2}) + \\ (\frac{5719}{36} v_1 - \frac{301}{18} v_2 + \frac{43}{36} v_3)(\vec{S}_i \circ \vec{S}_{i+2})^2 + \\ (-\frac{665}{18} v_1 + \frac{35}{9} v_2 - \frac{5}{16} v_3) (\vec{S}_i \circ \vec{S}_{i+2})^3 + \\ (-\frac{133}{12} v_1 + \frac{7}{6} v_2 - \frac{1}{12} v_3) (\vec{S}_i \circ \vec{S}_{i+2})^4
\end{multline*}

$\bullet \phantom{A} j = \frac{3}{2}$, $\epsilon = 0$:\\
\begin{multline*}
H = \sum_i (580 v_1) - 80 v_2 + 10 v_3 - 330 v_4 + 30 v_5)(\vec{S}_i \circ \vec{S}_{i + 1}) + \\ (91 v_1 - 11 v_2 2 v_3 - 91 v_4 11 v_5) (\vec{S}_i \circ \vec{S}_{i + 1})^2 + \\  (v_2 + v_5) (\vec{S}_i \circ \vec{S}_{i + 1})^3 + (v_1 + v_4) (\vec{S}_i \circ \vec{S}_{i + 1})^4 + \\ \frac{1}{6}(2275 v_1 - 275 v_2 + 25 v_3) (\vec{S}_i \circ \vec{S}_{i+2}) + \\ \frac{1}{36} (455 v_1 - 55 v_2 + 5 v_3)(\vec{S}_i \circ \vec{S}_{i+2})^2 + \\ \frac{1}{18} (-455 v_1 + 55 v_2 - 5 v_3) (\vec{S}_i \circ \vec{S}_{i+2})^3 + \\ \frac{1}{36} (-91 v_1 + 11 v_2 - v_3) (\vec{S}_i \circ \vec{S}_{i+2})^4
\end{multline*}

$\bullet$ No solutions found (with $n\leq 4$) for $ j = 2, \frac{5}{2}$.
\end{subsection}

\begin{subsection}{Spin $J = 3$}

Solutions (mostly cumbersome ones) were found for $j=1 (n=3)$, $j=2 (n=2)$ and $j=5/2 (n=2)$.
\end{subsection}
\end{section}
\newpage
\end{document}